\documentclass[aps,prx,twocolumn,showpacs,preprintnumbers,superscriptaddress,floatfix,amsmath,amssymb,amsfonts]{revtex4-1}
\usepackage{amsmath,amsfonts,amssymb}
\usepackage{color}
\usepackage{graphicx}
\usepackage[bookmarks=true,colorlinks,linkcolor=blue,urlcolor=blue,citecolor=blue]{hyperref}
\usepackage[center,sf,tight]{subfigure}
\usepackage[low-sup]{subdepth}
\usepackage{fouriernc}

\makeatletter
\@tfor\next:=abcdefghijklmnopqrstuvwxyzABCDEFGHIJKLMNOPQRSTUVWXYZ\do{%
  \def\command@factory#1{%
    \expandafter\def\csname bf#1\endcsname{\mathbf{#1}}
  }
 \expandafter\command@factory\next
}
\@tfor\next:=abcdefghijklmnopqrstuvwxyzABCDEFGHIJKLMNOPQRSTUVWXYZ\do{%
  \def\command@factory#1{%
    \expandafter\def\csname bb#1\endcsname{\mathbb{#1}}
  }
 \expandafter\command@factory\next
}
\@tfor\next:=abcdefghijklmnopqrstuvwxyzABCDEFGHIJKLMNOPQRSTUVWXYZ\do{%
  \def\command@factory#1{%
    \expandafter\def\csname cal#1\endcsname{\mathcal{#1}}
  }
 \expandafter\command@factory\next
}
\@tfor\next:=abcdefghijklmnopqrstuvwxyzABCDEFGHIJKLMNOPQRSTUVWXYZ\do{%
  \def\command@factory#1{%
    \expandafter\def\csname til#1\endcsname{\tilde{#1}}
  }
 \expandafter\command@factory\next
}
\makeatother

\newcommand{\up}{\uparrow}
\newcommand{\dn}{\downarrow}

\newcommand{\bracket}[1]{\left\langle#1\right\rangle}

\begin{document}
\title{Cold-spots and  glassy nematicity  in underdoped cuprates}
\author{Kyungmin Lee}
\affiliation{Department of Physics, Cornell University, Ithaca, NY 14853, USA}
\author{Steven A. Kivelson}
\affiliation{Department of Physics, Stanford University, Stanford, CA 94305, USA}
\author{Eun-Ah Kim}
\affiliation{Department of Physics, Cornell University, Ithaca, NY 14853, USA}

\begin{abstract}
There is now copious direct experimental evidence of  various forms of (short-range) charge order in  underdoped cuprate high temperature superconductors, and spectroscopic signatures of a nodal-antinodal dichotomy in the structure of the single-particle spectral functions.  In this context, we analyze the Bogoliubov quasiparticle spectrum in a superconducting nematic glass.  The coincidence of  the superconducting ``nodal points'' and the nematic  ``cold-spots''  on the Fermi surface naturally accounts for many of the most salient features of the measured spectral functions (from angle-resolved photoemission) and the local density of states (from scanning tunneling microscopy).
\end{abstract}

\maketitle

\section{Introduction}

The existence of glassy charge order in the pseudogap phase of cuprates is now well established:
Both momentum space and real space probes find charge-density-wave (CDW) order with moderate (but never infinite) correlation lengths~\cite{tranquada_evidence_1995,zimmermann_hardxray_1998,abbamonte_spatially_2005,ghiringhelli_longrange_2012,blackburn_xray_2013,comin_charge_2014,neto_ubiquitous_2014,fujita_direct_2014,chang_magnetic_2016,gerber_threedimensional_2015,[{For a recent review, see }][]fradkin_colloquium_2015}.
Evidence of a tendency to nematic order has been adduced from local probes~\cite{howald_periodic_2003,lawler_intraunitcell_2010,wu_incipient_2015}, diffraction~\cite{hinkov_electronic_2008,achkar_nematicity_2016}, and transport~\cite{ando_electrical_2002,daou_broken_2010}.
Much of the associated theory literature has focused on either uniform long-range ordered states, or dynamically fluctuating order.
In contrast, glassy order implies strong static heterogeneities, which complicate any theoretical analysis.

The basic superconducting state is thought to be reasonably well described by a simple mean-field theory with a $d$-wave superconducting gap.
Nevertheless, when the superconductivity coexists with glassy charge order, spectroscopic measurements reveal a number of ``anomalous'' features that are not simply related to any long-range order.
It is thus worth asking whether some or all of these anomalous features are a consequence of glassy charge order.
Heterogeneous order parameters have  been studied previously in the context of cuprates~\cite{carlson_hysteresis_2006,robertson_distinguishing_2006,delmaestro_stripe_2006,kim_theory_2008,kaul_imaging_2008,phillabaum_spatial_2012,nie_quenched_2014,russo_random_2016}. 
But most of these works have focused on the effects of quenched randomness (e.g., impurities) on the ordering tendencies themselves.
Here instead we study how the heterogeneity associated with glassy order affects various spectroscopic properties.

Technically, our approach is similar to that employed in earlier works on the effects of point-like impurities~\cite{
trivedi_superconductorinsulator_1996,wang_quasiparticle_2003,garg_strong_2008,kim_interference_2010,bouadim_single_2011,balatsky_impurityinduced_2006}.
However, because the glassy order is assumed to reflect (in part)  the system's tendency toward symmetry breaking, in the present study the effective scattering (``disorder'') potential is taken to have two properties not present in earlier studies: (1)  a moderate correlation length, and (2) a non-trivial form factor.
Although we do consider various forms of CDW order, our most extensive and most significant results are associated with a nematic glass, which by symmetry has a $d$-wave form factor. 
While the lack of translation symmetry destroys the long range coherence of  the quasiparticles, the $d$-wave form factor gives rise~\cite{metzner_soft_2003,yamase_fermi_2012,okamoto_spontaneous_2012} to cold-spots~\cite{ioffe_zonediagonaldominated_1998}, near which the quasiparticles are increasingly weakly coupled to the glassy order.  Because these cold-spots coincide with the nodal points in a $d$-wave superconductor, the lowest energy quasiparticles are also the most weakly affected by the nematic glass.

\begin{figure}
\subfigure[][]{\raisebox{9mm}{\includegraphics[height=0.8in]{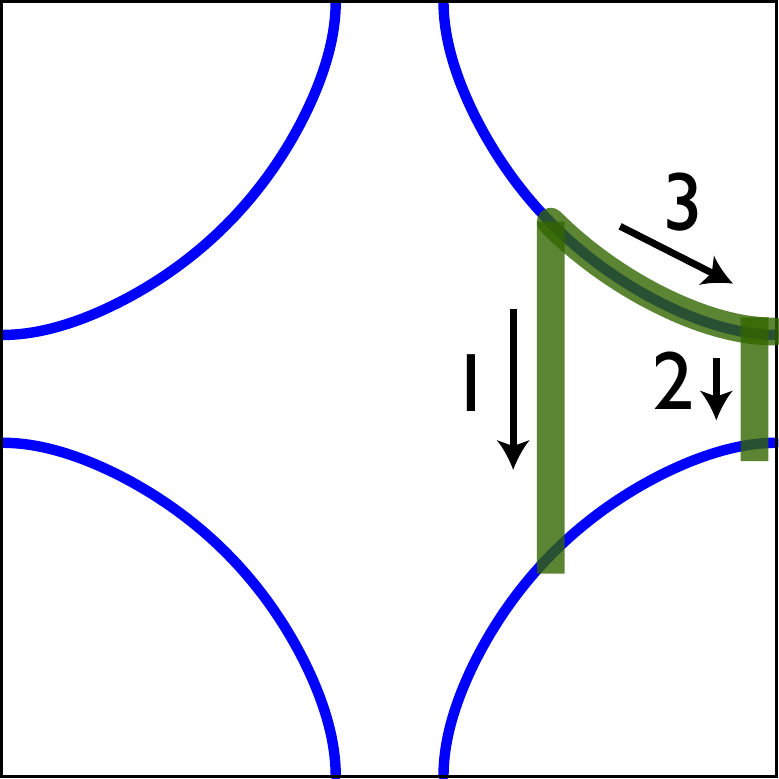}}}
\;
\subfigure[][]{\includegraphics[]{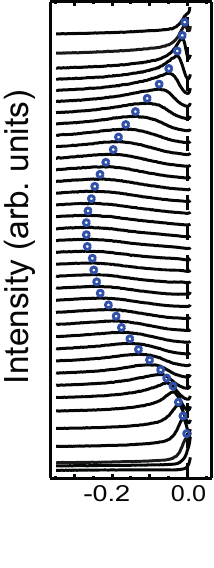}}\;\;
\subfigure[][]{\includegraphics[]{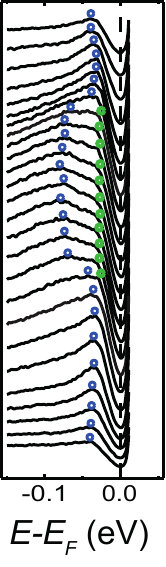}}\;
\subfigure[][]{\includegraphics[]{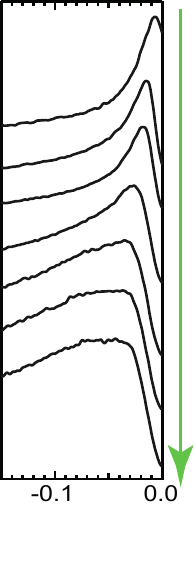}}
\caption{\label{fig:ARPES}
Energy distribution curves (EDCs) from ARPES measurements reproduced from \textcite{he_singleband_2011}(Bi-2201).
(b)--(d) Different paths in $\bfk$-space as shown in (a): (b) is a ``nodal'' cut through the Fermi surface [path 1 of (a)], (c) is an antinodal cut through the Fermi surface [path 2 of (a)], and (d) is a path along the Fermi surface starting at the nodal point and ending at the antinodal [path 3 of (a)].
The measured spectral weight has been divided by the Fermi function.
}
\end{figure}

In comparing our results to experiment,  we consider features from three different experiments: angle-resolved photomemission spectroscopy (ARPES), scanning tunneling microscopy (STM), and optical measurements:

(1) The most salient feature of ARPES that we address is the ``nodal-antinodal'' dichotomy.  The energy distribution curves (EDCs) for momenta along a cut across the Fermi surface [Fig.~\ref{fig:ARPES}(b)] in the nodal region consist of a single dispersing feature which at least roughly resembles that expected of a quasiparticle with a finite lifetime.  Conversely, along a similar cut perpendicular to the antinodal segment of the Fermi surface [Fig.~\ref{fig:ARPES}(c)], the EDC is complex, exhibiting at least two distinct features with  apparent dispersion relations (if that notion applies at all) that appear almost discontinuous.
Nevertheless, moving along the Fermi surface from the nodal to the antinodal point, the EDC curves evolve smoothly and  monotonically [Fig.~\ref{fig:ARPES}(d)] with no sign of any sharp boundary, or of the non-monotonic behavior one would expect if there were ``hot-spots'' on the Fermi surface corresponding to the spanning vectors associated with incipient density-wave order.

(2) Much thought has gone into the analysis of the rich structural and spectroscopic information encoded in the variations of the local density of states (LDOS) measured by STM, especially on BSCCO.
Here we focus exclusively on a  clear ``dichotomy'' [Fig.~\ref{fig:LDOS}(a)] that has been apparent since the earliest studies~\cite{howald_inherent_2001,pan_microscopic_2001,pasupathy_electronic_2008}:
At relatively low energies, the LDOS is remarkably homogeneous and has the V-shaped energy dependence expected for a uniform $d$-wave superconductor, while at energies comparable to the gap (or pseudo-gap), there are order one variations of the LDOS as a function of position.
Note that the ``dispersing features'' in the Fourier transform of the LDOS which have been identified with quasiparticle interference effects are more or less confined to the ``low energy'' range in which the LDOS is relatively homogeneous.

(3) The low $T$ optical conductivity rises  roughly linearly with increasing frequency $\omega$ to a peak at $\omega\sim 100-200$meV that
(at least in the more recent data on Hg-1201)  is larger than any reasonable estimate of the superconducting gap, and then drops slowly at larger $\omega$ [Fig.~\ref{fig:opticon}(a) and \ref{fig:opticon}(b)].
All of these features are somewhat anomalous, as is the $T$ dependence of $\sigma(\omega)$.

As we shall show, these salient features of the ARPES and STM experiments are naturally explained by the coincidence of the nematic cold-spots and the superconducting nodes in a superconducting nematic glass.  This is illustrated in Figs.~\ref{fig:ARPESTheory} and \ref{fig:LDOS}(b), respectively.
We also find that the optical conductivity computed in the simplest model of such a glass looks remarkably like the experiments [Fig.~\ref{fig:opticon}(c)].  However,  concerning the thermal evolution of $\sigma$, there are aspects of the solution that are slightly problematic, since in making the comparison at the higher temperatures, we are comparing experimental results at $T>T_c$ with theoretical results at $T< T_c$.

The notion that various phenomena in the cuprates may be associated with the existence of cold-spots on the Fermi surface is not new.  Notably, a number of earlier studies~\cite{hussey_angular_1996,ioffe_zonediagonaldominated_1998,zheleznyak_phenomenological_1998,abdel-jawad_anisotropic_2006,abdel-jawad_correlation_2007,hussey_phenomenology_2008,french_tracking_2009} have suggested  that salient features of the  transport properties of the ``normal'' (``bad metal'' or ``strange metal'') state can be interpreted as evidence of a strongly anisotropic scattering rate on the Fermi surface, with cold-spots along the zone diagonal. 
Anisotropic scattering rates inferred from ARPES data supports the case~\cite{kaminski_momentum_2005}.
In contrast, in the present study,  the  focus is primarily on the low temperature properties of the system where superconductivity and pseudo-gap signatures coexist.  In this regime, glassy nematicity provides a plausible microscopic origin of  anisotropic scattering rates.  To the extent that there is a relation to the cold-spots of the earlier proposals, it is more likely that at higher doping and larger temperature they are associated with quantum or thermal nematic fluctuations~\cite{dellanna_electrical_2007,schattner_ising_2015,lederer_unpublished_}, rather than with frozen, glassy nematic order.

\section{The Model}
As our primary focus is on the quasiparticle properties deep in the superconducting state and far from any quantum phase transition, where neither thermal nor quantum fluctuations are expected to be significant, we assume that it is sufficient to study the solutions of an appropriate mean-field Bogoliubov-de Gennes Hamiltonian, 
\begin{align}
  H_{\mathrm{BdG}}
    &= \sum_{\bfx \bfy}
        \begin{pmatrix}c_{\bfx\up}^\dag &  c_{\bfx\dn} \end{pmatrix}
        \begin{pmatrix}
          t_{\bfx \bfy}          & \Delta_{\bfx \bfy} \\
          \Delta_{\bfy \bfx}^{*} & -t_{\bfy \bfx}
        \end{pmatrix}
        \begin{pmatrix}c_{\bfy\up} \\ c_{\bfy\dn}^\dag \end{pmatrix},
\label{eq:hambdg}
\end{align}
where $c_{\bfx\sigma}$ annihilates an electron at site $\bfx$ with spin $\sigma$.  
The ``normal'' part of the Hamiltonian is assumed to be of the form
\begin{align}
 t_{\bfx \bfy} = t^{(0)}_{\bfx \bfy} + V(\bfx,\bfy),
 \end{align}
 where the first term represents the underlying band-structure, $t^{(0)}_{\bfx\bfy} = 
-\mu \delta_{\bfx \bfy}
-t \delta_{\langle  \bfx,  \bfy \rangle }
-t' \delta_{\langle\langle \bfx,  \bfy \rangle\rangle}$,
with $t = 1$, $t' = -0.3$, and chemical potential $\mu=-0.8$, and the term $V$ represents the effective potential due to  the presence of (glassy) charge order.
The anomalous term $\Delta_{\bfx\bfy}$ on each pair of nearest-neighbor sites $\langle \bfx, \bfy \rangle$ is determined self-consistently from the gap equation
\begin{align}
\label{eq:BCS}
\Delta_{\bfx \bfy}
  &= 
    \frac{U}{2}
    \bracket{c_{\bfy\up} c_{\bfx\dn} + c_{\bfx\up} c_{\bfy\dn}},
\end{align}
with $\Delta_{\bfx \bfy}=0$ otherwise.
A value of $U=0.732t$  is chosen so that in the clean limit [$V(\bfx,\bfy)=0$], the transition to the $d$-wave superconducting state occurs at $T_{c}^{0} = 0.05t$, and the resulting uniform $d$-wave BCS ground state has a gap-function in $\bfk$ space: $\Delta_{\bfk} = \Delta_0 (\cos k_x - \cos k_y)$ with $\Delta_0=0.055t$.
[Note that this unrealistic pairing strength was chosen such that 
the antinodal gap ($\sim 0.1 t$) is larger than the energy resolution set by $v_F^{\text{max}} \Delta k \sim 0.06t$, where $v_{F}^{\text{max}}$ is the maximum Fermi velocity, and $\Delta k = 2 \pi / N$ is the momentum resolution, for system size $N = 256$ used in most calculations, and yet reasonably smaller than the energy difference between the Fermi level and van Hove singularity ($\Delta E_{\text{vH}} = 0.4t$).]

Finally,  the effect of any (glassy) charge order is represented by  a local order parameter,  $\varphi(\bfx)$, (taken to be real under the assumption that  time-reversal symmetry is unbroken), which couples to the quasiparticles with a ``form factor'' $f(\bfr)$:
\begin{align}
\label{eq:coupling}
  V(\bfx,\bfy) 
    &=
      \frac{1}{2} f(\bfx - \bfy) 
        \left[ \varphi(\bfx) + \varphi(\bfy) \right].
\end{align}
For typical random disorder, or for the simplest forms of charge-density-wave (CDW) order, $f(\bfr) \approx \delta_{\bfr,\boldsymbol{0}}$, corresponding to a position-dependent single-site energy.  
In contrast, for nematic order, $f(\bfr)$ must flip sign under 90$^{\circ}$ rotation by symmetry; to be explicit we choose the shortest-range form factor compatible with nematic symmetry, $f(\bfr) = \delta_{\bfr, \pm\hat{x}} - \delta_{\bfr,\pm\hat{y}}$, corresponding to a position dependent modulation of the nearest-neighbor hopping matrix elements.
We assume $\varphi(\bfx)$ are random variables chosen from an ensemble defined by the configuration average of the two-point correlator, 
$\overline{\varphi(\bfx) \varphi(\bfx + \bfr)} = \Gamma(\bfr)$.
The spatial range of the assumed correlations, as well as any tendency to ordering with non-zero period (as in a CDW with a finite ordering vector $\bfQ$) are encoded in $\Gamma(\bfr)$.
In the case of a nematic glass, we take $\Gamma(\bfr) =\Gamma_{\mathrm{nem}} \exp ( - r^2/2 \xi_\mathrm{nem}^2 )$, where $\Gamma_{\mathrm{nem}}$ is a measure of the mean-square magnitude of the nematic order, and $\xi_\mathrm{nem}$ is the nematic correlation length.
For a CDW glass, $\Gamma(\bfr) =(\Gamma_{\mathrm{cdw}}/2) [\cos({\bf Q}\cdot\bfr) +\cos({\bf Q}^\prime\cdot\bfr)]\exp ( - r^2/2 \xi_\mathrm{cdw}^2 )$ where ${\bf Q}$ and ${\bf Q}^\prime$ are the two symmetry related ordering vectors.

We can already see how a glassy nematic will generate 
cold-spots by simply Fourier transforming Eq.~\eqref{eq:coupling} to yield
\begin{align}
\label{eq:couplingmomentum}
  \tilde{V}(\bfk,\bfp)
  &=
      \frac{1}{2}\left[ \tilde{f}(\bfk) + \tilde{f}(\bfp) \right]
      \tilde{\varphi}(\bfk - \bfp) ,
\end{align}
where $\tilde f(\bfk)$ and $\tilde \varphi(\bfq)$ are respective Fourier transforms of $f(\bfr)$ and $\varphi(\bfx)$.
For nematic order, $\tilde f(\bfk) = 2 ( \cos k_x - \cos k_y )$.
When $\varphi(\bfx)$ is uniform, $\tilde{\varphi}(\bfq)$ is a delta function peaked at $\bfq = 0$, in which case $\tilde{V}$ simply leads to distortion of Fermi surface [dashed line in Fig.~\ref{fig:spectral}(a)].
When $\varphi(\bfx)$ is non-uniform, on the other hand, $\tilde{\varphi}(\bfq)$ is no longer a delta function, and momentum states acquire lifetimes by scattering off of $\tilde{\varphi}(\bfq)$.
The form factor $\tilde{f}(\bfk)$ gives rise to strong anisotropy of the quasiparticle lifetimes: While the antinodal quasiparticles are strongly scattered, the ``nodal quasiparticles'' at the cold-spots ($|k_x|=|k_y|$) are largely unaffected.
The cold-spots arise solely as a result of the symmetry of the (local) nematic order.

\section{Method of Solution}

To achieve sufficiently fine $\bfk$-space resolution for present purposes, we work with a system with periodic boundary conditions of size $N\times N$ with $N=128$ or where needed $N=256$ or $512$.
However, because it is computationally intensive to solve the self-consistency equations for such a large system, we have chosen the disorder potential $V(\bfx,\bfy)$ (and correspondingly $\Delta_{\bfx \bfy}$) to be periodically repeated in blocks of size $L\times L$ with $L=32$.
This compromise allows us to study real-space heterogeneity, while at the same time reducing the finite size effect through fine $\bfk$-space (and hence energy) resolution.

We generate a configuration of the quenched variables by choosing $\{\varphi(\bfx)\}$ from a distribution with a Gaussian two-point correlator $\Gamma(\bfr)$.
To avoid long-range correlation of $\varphi(\bfx)$, we choose $\xi_{\mathrm{nem}}$ to be small relative to the size of the repeated blocks.
Specifically, we require $\vert \Gamma(\bfr)/\Gamma(0) \vert < 1\%$ at $|\bfr|=La/2$, where $a$ is the lattice constant, %we 
which means we are limited to $\xi_{\mathrm{nem}} \le 5a$.
To be concrete, we will present results primarily for $\xi_{\mathrm{nem}}=2a$;
although this is shorter than typical correlation lengths of glassy order measured in experiments (as defined in Ref.~\cite{robertson_distinguishing_2006} for example), we chose it for two reasons:
(1) As we will find in our spectral function analyses, $\xi_{\mathrm{nem}}=2a$ results show qualitatively no difference with $\xi_{\mathrm{nem}}=4a$.
(2) Obviously, results for short correlation lengths suffer less from finite size effect than longer correlation lengths.
For each configuration of $\{ \varphi(\bfx) \}$, we determine the values of $\Delta_{\bfx \bfy}$ from the solution of the self-consistency equation Eq.~\eqref{eq:BCS}.
For example, a typical configuration of $\varphi(\bfx)$ is shown in Fig.~\ref{fig:realspace}(a) generated from an ensemble with $\sqrt{\Gamma_{\mathrm{nem}}}=0.1t$ and $\xi_{\mathrm{nem}}=2a$;  the corresponding self-consistently determined gap function $\Delta_{\bfx\bfy}$ is shown in Fig.~\ref{fig:realspace}(b).
While there are clearly significant variations in the magnitude of the pair-fields from place to place, the $d$-wave character of the sign structure is universally preserved;  it is positive on all $x$-directed and negative on all $y$-directed bonds.

Finally, once self-consistency is achieved, we calculate three spectroscopic observables: (1) the ARPES spectral function $A(\bfk, E)$, (2) the local density of states $n(\bfx, E)$, and (3) the optical conductivity $\sigma(\omega)$.
The spectroscopic observables we study are self-averaging properties.
Although here we present results from a single configuration, we have confirmed that different configurations of $\{\varphi(\bfx)\}$ generated probabilistically from the same distribution result in minor quantitative changes in the calculated spectra, with no significant qualitative difference.

\section{Results for the superconducting nematic glass}

\begin{figure}
\subfigure[][]{\includegraphics[height=2.4in]{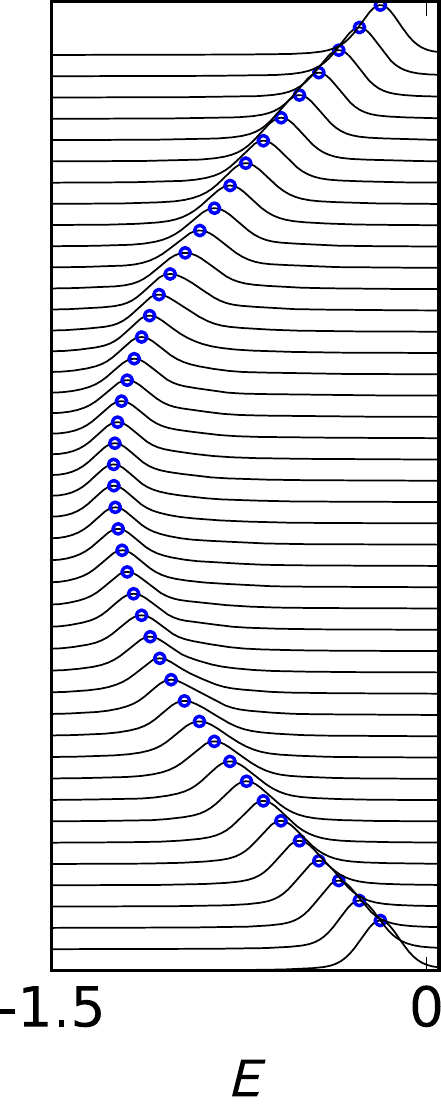}}\quad\;
\subfigure[][]{\includegraphics[height=2.4in]{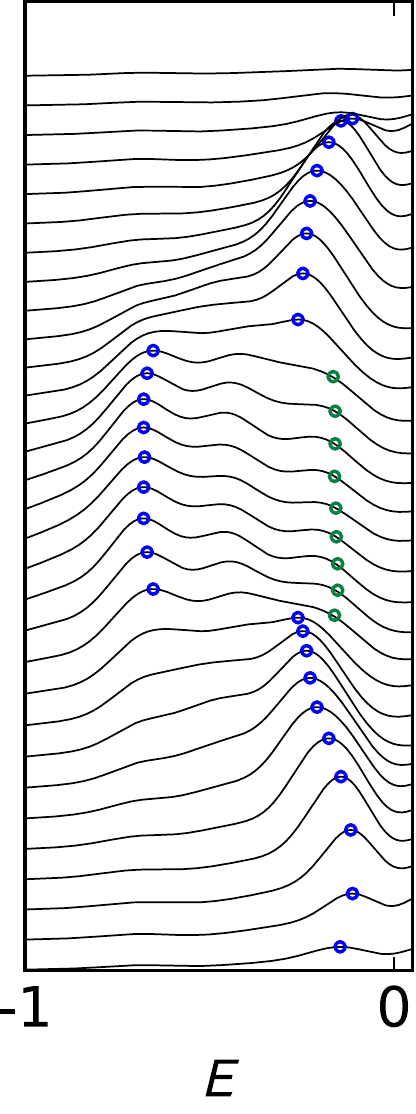}}\quad
\subfigure[][]{\includegraphics[height=2.4in]{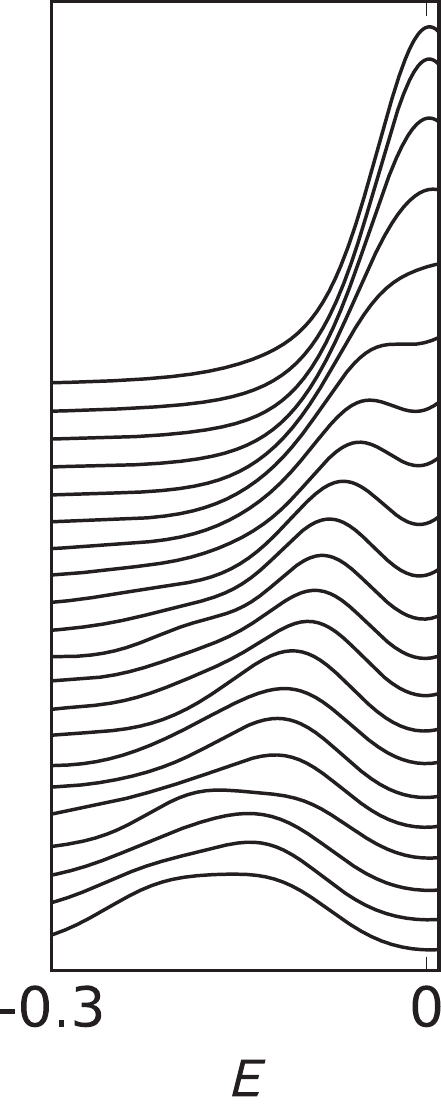}}
\caption{\label{fig:ARPESTheory}
EDC's computed along the same paths in $\bfk$-space as shown in  Fig. 1a for a superconducting nematic glass with $\Delta_0=0.055t$, $\sqrt{\Gamma_{\mathrm{nem}}} = 0.1 t$, and $\xi_{\mathrm{nem}}=2a$:
(a) is a ``nodal'' cut through the Fermi surface [path 1 of Fig.~\ref{fig:ARPES}(a)], and
(b) is an antinodal cut through the Fermi surface [path 2 of Fig.~\ref{fig:ARPES}(a)].
A blue circle marks the maximum of each curve, and a green circle marks the position of a ``shoulder''.
(c) is a path along the Fermi surface starting at the nodal point and ending at the antinodal point [path 3 of Fig.~\ref{fig:ARPES}(a)].
}
\end{figure}

\begin{figure}
\quad
\subfigure[][]{\includegraphics[height=1.3in]{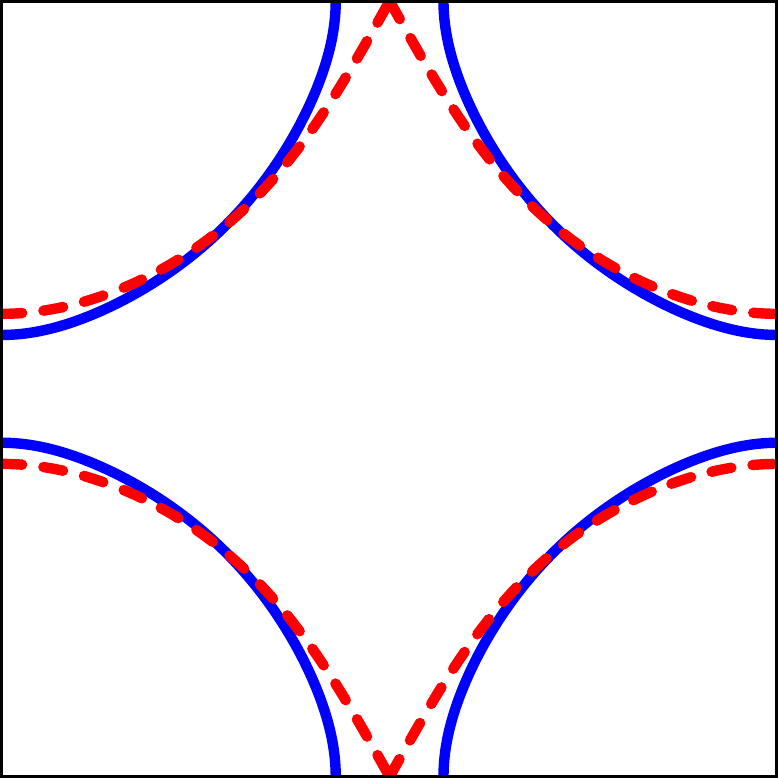}}\qquad
\subfigure[][]{\includegraphics[height=1.3in]{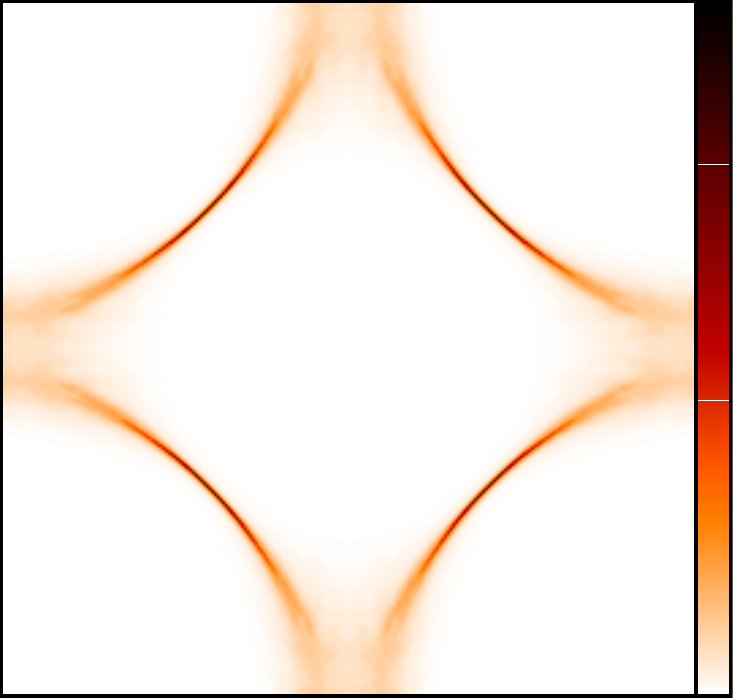}}\\
\subfigure[\label{fig:selfenergy-simulation-nematic}]{\includegraphics[height=1.5in]{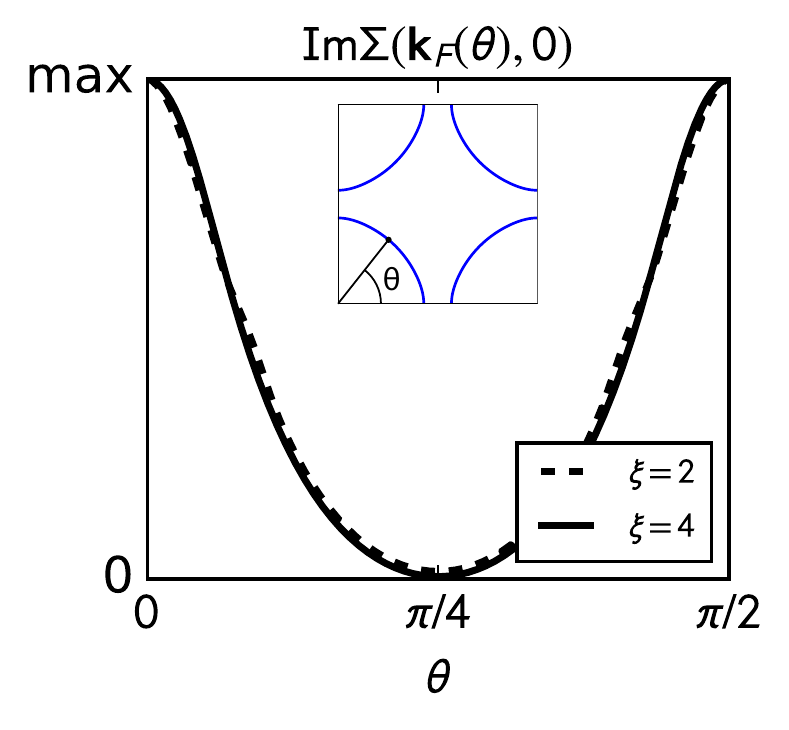}}%
\subfigure[\label{fig:selfenergy-nematic}]{\includegraphics[height=1.5in]{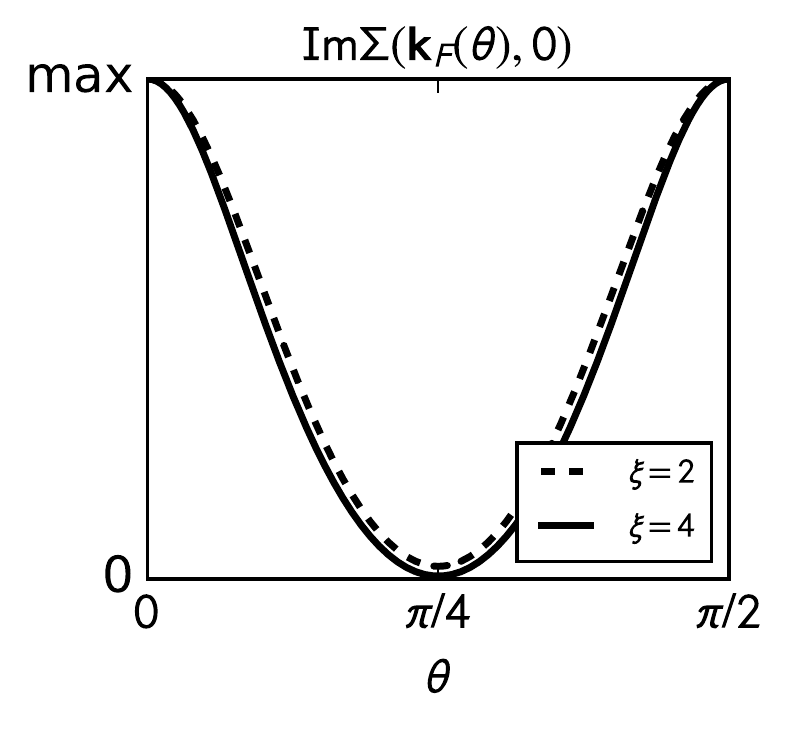}}
\\
\caption{\label{fig:spectral}
(a) Fermi surface of the model system in the normal state ($\Delta_{\bfx\bfy}=0$). Solid lines are computed in the symmetric phase ($\varphi(\bfx)=0$) and the dashed lines in a uniform nematic phase ($\varphi(\bfx)=0.05t$.)
(b) $A(\bfk,E)$ of the superconducting nematic glass at a fixed energy $| E=-0.2t| >  2\Delta_0 = 0.11t$ showing the nodal antinodal dichotomy.  Color intensity indicates the magnitude.
(c) Imaginary part of normal state electronic $\omega=0$ self-energy on the Fermi surface coupled to nematic order extracted from real space simulation as a function of angle around the Fermi surface $\theta$ as defined in the inset, and (d) calculated in the Born approximation.
}
\end{figure}

\begin{figure}
\subfigure[][]{\includegraphics[height=1.35in]{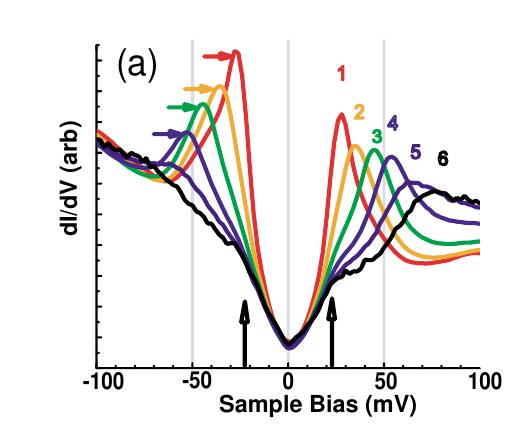}}
\subfigure[][]{\includegraphics[height=1.35in]{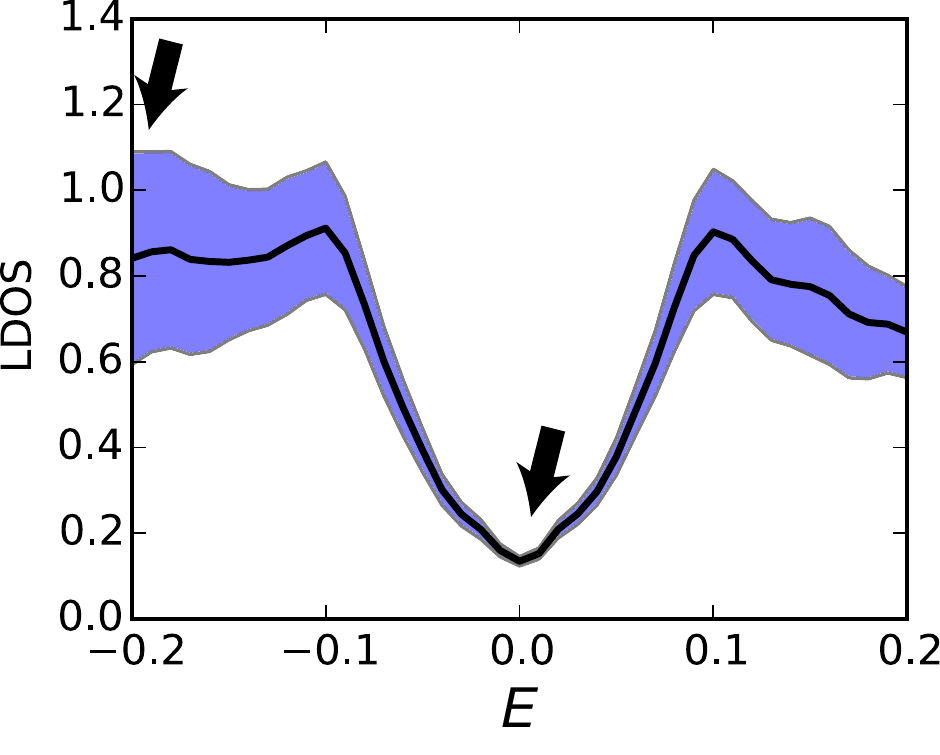}}
\caption{\label{fig:LDOS}
Local density of states as a function of energy measured at multiple locations on the surface:
(a) Results of STM measurements on Bi-2212 reproduced from Ref.~\onlinecite{mcelroy_coincidence_2005}.
Different curves represent tunneling spectra measured at different locations of the sample.
(b) Computed for a superconducting nematic glass with $\Delta_0=0.055t$, $\sqrt{\Gamma_{\mathrm{nem}}} = 0.1 t$, and $\xi_{\mathrm{nem}}=2a$;
the solid black curve and the shaded region indicate the spatially averaged DOS and spatial standard deviation of LDOS, respectively.
The spatial maps of LDOS at energies marked by the two arrows are shown in Figs.~\ref{fig:realspace}(c) and \ref{fig:realspace}(d).
}
\end{figure}

\begin{figure}\begin{center}
\subfigure[][]{\includegraphics[height=1.3in]{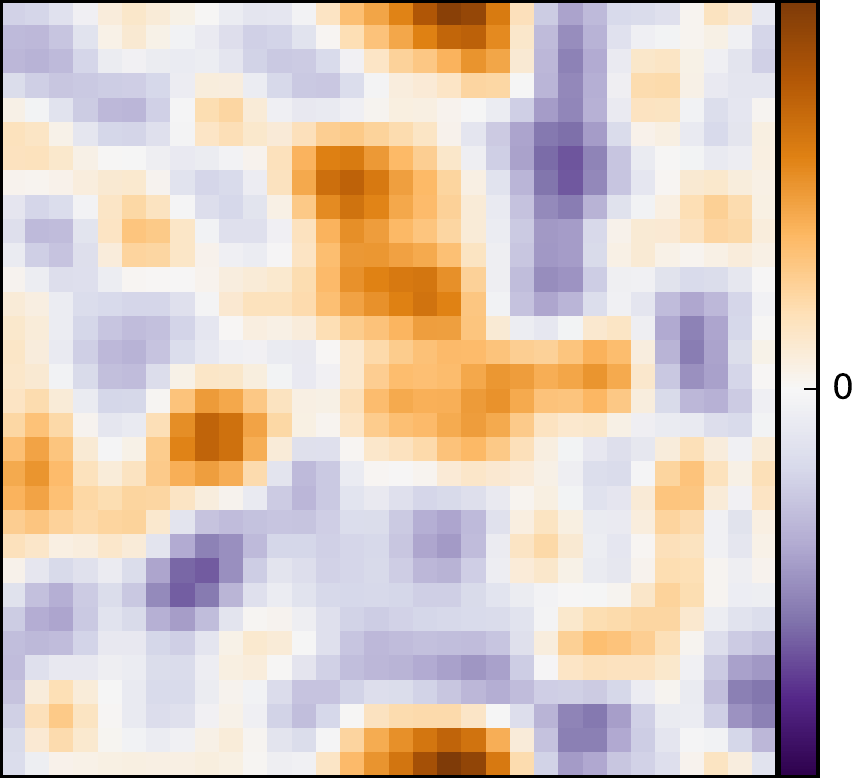}}\quad
\subfigure[][]{\includegraphics[height=1.3in]{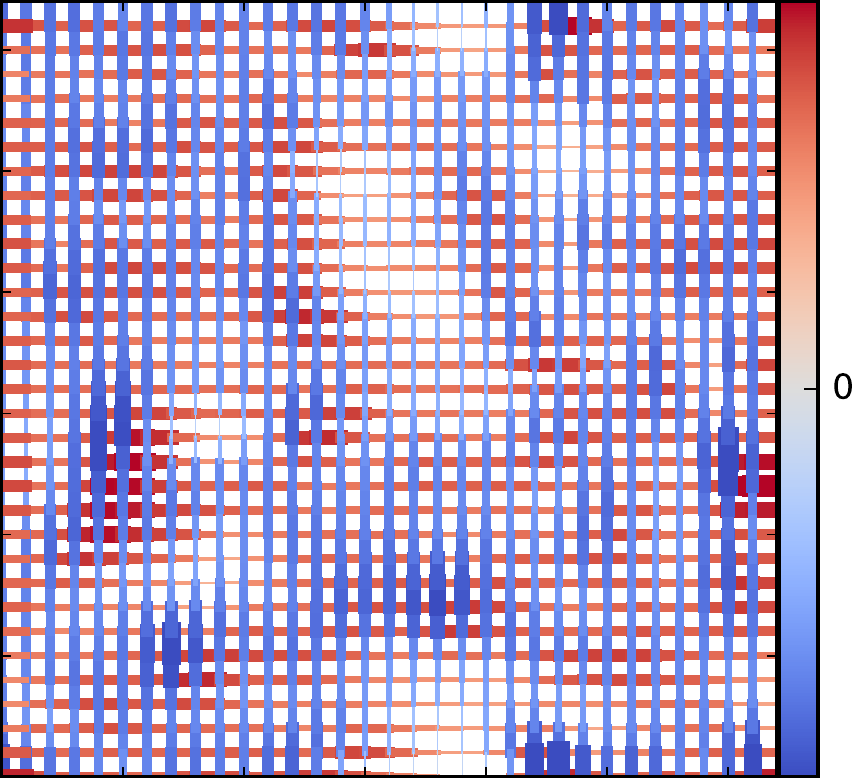}}\\
\quad
\subfigure[][]{\includegraphics[height=1.45in]{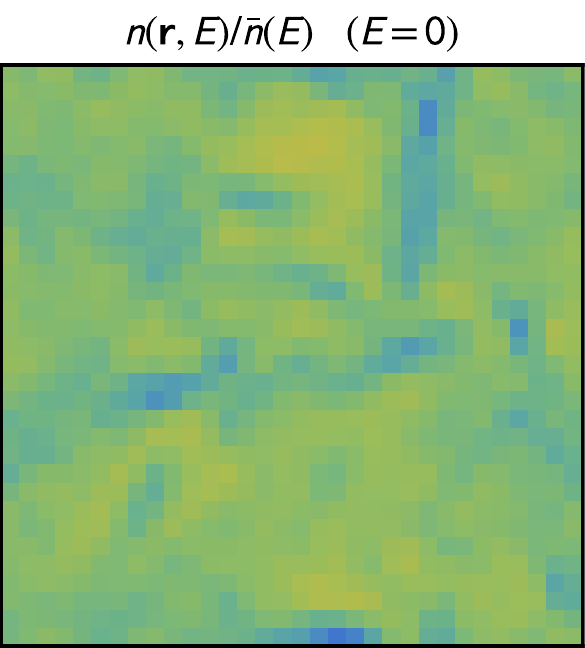}}\;\;\quad
\subfigure[][]{\includegraphics[height=1.45in]{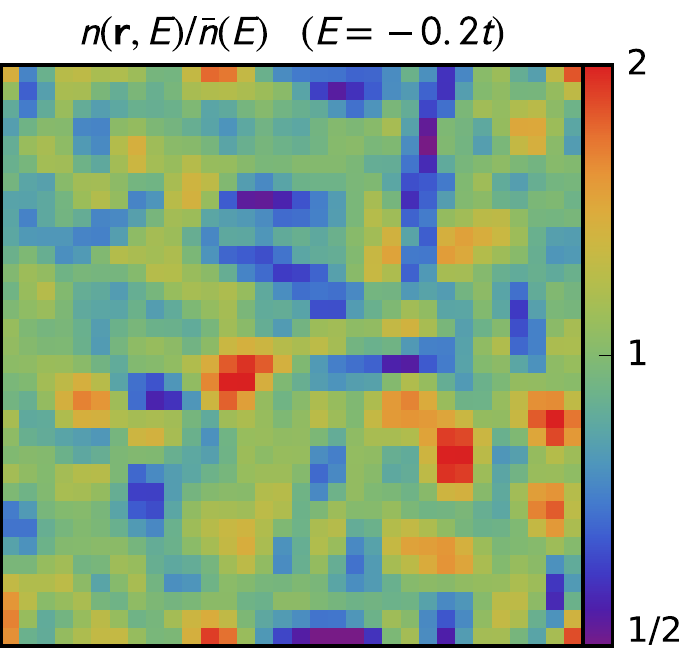}}
\caption{\label{fig:realspace}
(a) A representative configuration of $\varphi(\bfx)$ representing a nematic glass with $\xi_{\mathrm{nem}}=2a$ in a system of size $32\times 32$ unit cells.
(b) The gap parameter, $\Delta_{\bfx\bfy}$, at $T=0$ determined self-consistently with $\varphi(\bfx)$ shown in Fig.~\ref{fig:realspace}(a) when the root-mean-square magnitude $\sqrt{\Gamma_{\mathrm{nem}}}=0.1t$.
The sign of $\Delta_{\bfx \bfy}$ on each bond is represented by the color (red is positive blue is negative) with the magnitude represented by the thickness of the line as well as opacity.
Manifestly, the local symmetry of the pairing is uniformly $d$-wave.
The associated normalized LDOS $n(\bfx, E)/\bar{n}(E)$, is shown for (c) $E=0$  and (d) $E=-0.2t$.
}
\end{center}\end{figure}

\begin{figure}\begin{center}
\subfigure[]{\includegraphics[width=1.7in]{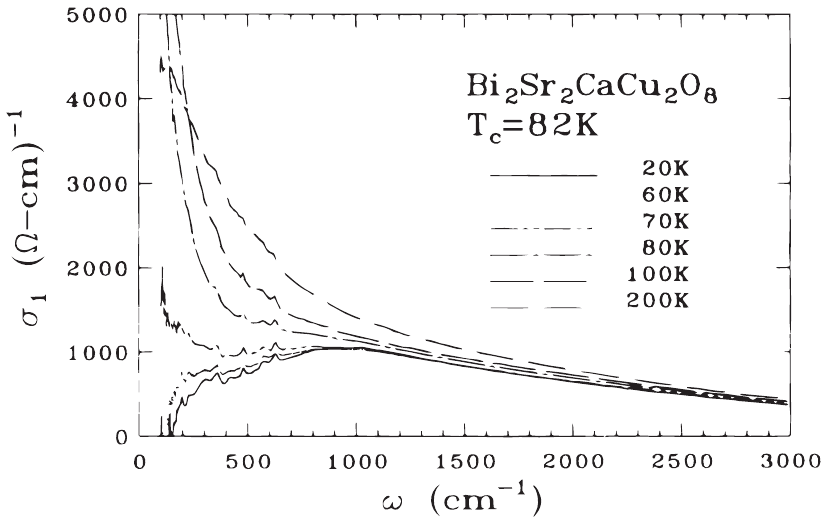}}
\subfigure[]{\includegraphics[width=1.45in]{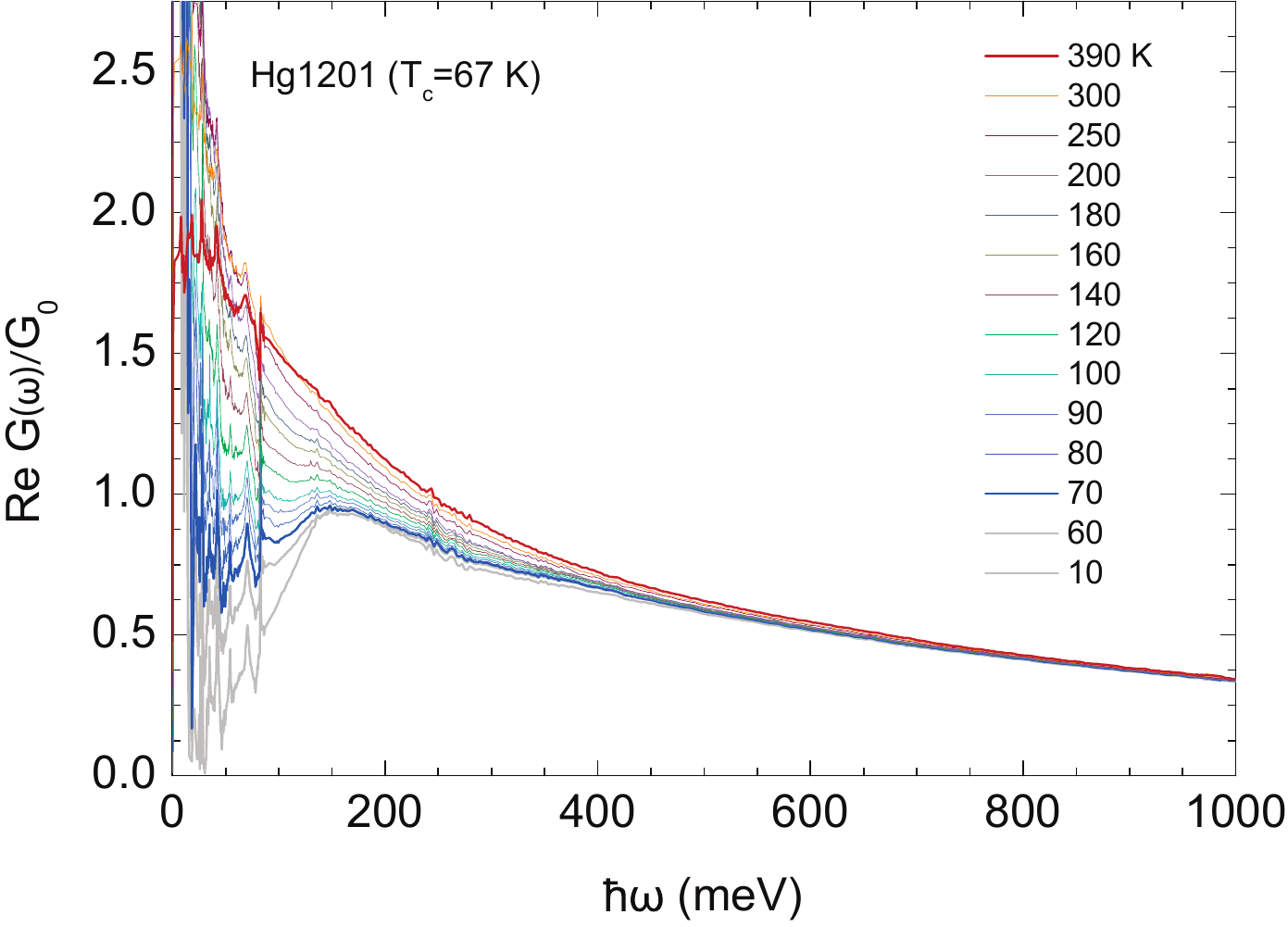}}\\
\subfigure[]{\includegraphics[height=1.3in]{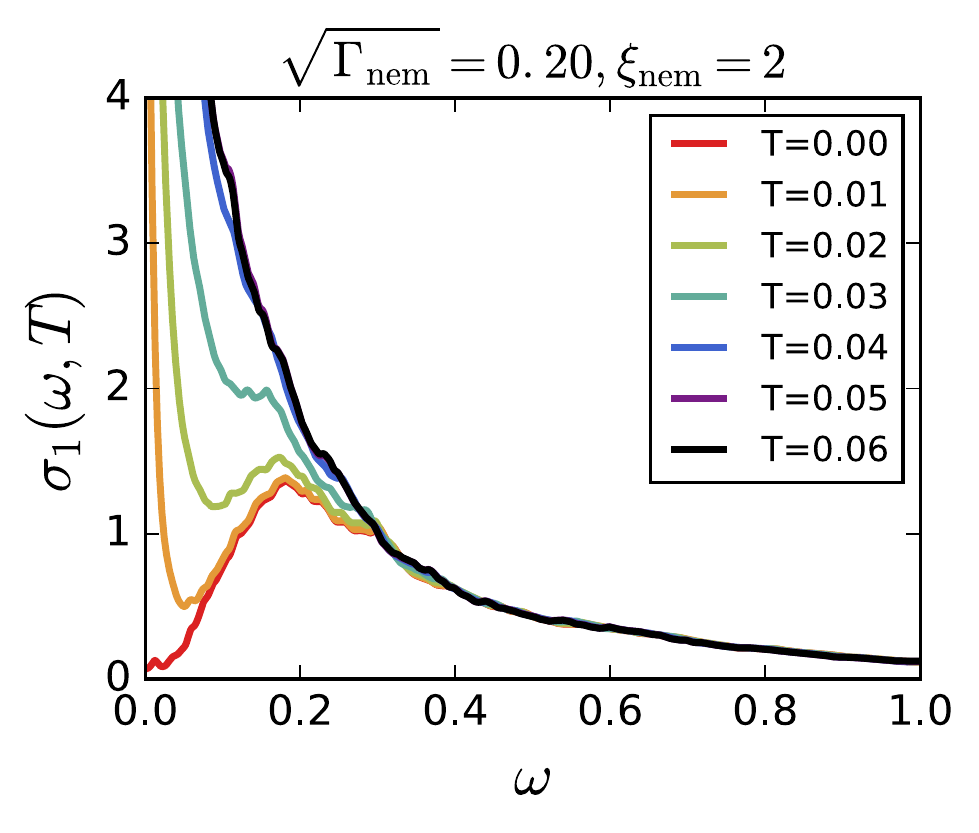}}\quad
\subfigure[]{\includegraphics[height=1.3in]{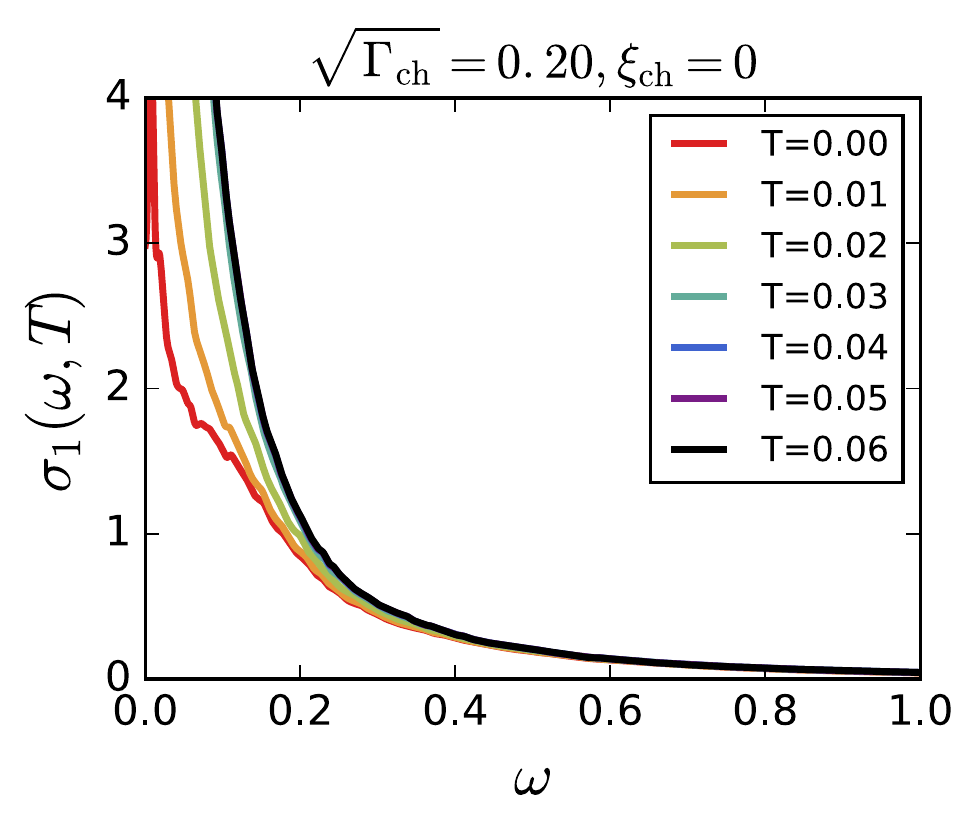}}
\caption{\label{fig:opticon}Optical conductivity
(a) from experiment (Ref.~\onlinecite{romero_quasiparticle_1992}) on Bi-2212 with $T_c=82$K, 
(b) from experiment (Ref.~\onlinecite{mirzaei_spectroscopic_2013}) on Hg-1201 with $T_c=67$K,
(c) calculated for the superconducting nematic glass with  $\xi_{\text{nem}}=2a$ and $\sqrt{\Gamma_{\text{nem}}}=0.2t$), and
(d) calculated for a disordered superconductor, where the disorder is assumed to have an on-site $s$-wave form factor  with $\xi_{\text{ch}}=0$ and $\sqrt{\Gamma_{\text{ch}}}=0.2t$.
In both (c) and (d), $\Delta_{\bfx \bfy}$ is self-consistently determined by Eq.~\ref{eq:BCS} with $U=0.732t$.
}
\end{center}\end{figure}

Among various forms of glassy charge order that we have considered, the nematic glass best reproduces the nodal-antinodal dichotomy as observed in ARPES (Fig.~\ref{fig:ARPES}).
We have carried out calculations for various choices of the strength and correlation length of the nematic order, but to be concrete we present representative data corresponding to $\sqrt{\Gamma_{\mathrm{nem}}}=0.1t$ and $\xi_{\mathrm{nem}} = 2a$.
In Fig.~\ref{fig:ARPESTheory} we show our results for the spectral function, such as would be measured in ARPES. 
In Fig.~\ref{fig:ARPESTheory}a, the EDCs (i.e. the energy dependence of $A(\bfk,E)$ at fixed $\bfk$) for $\bfk$'s along a cut through the nodal point on the Fermi surface shows a quasiparticle-like dispersion that is otherwise featureless, as in ARPES measurements [Fig.~\ref{fig:ARPES}(b)].
On the other hand, the EDCs in the antinodal region, shown in Fig.~\ref{fig:ARPESTheory}(b), have two branches which are almost discontinuous, a quasiparticle dispersion and a shoulder fixed at the superconducting gap scale, reminiscent of Fig.~\ref{fig:ARPES}(c).
EDCs along the Fermi surface from node to antinode [Fig.~\ref{fig:ARPESTheory}(c)] also qualitatively agree with the ARPES measurements [Fig.~\ref{fig:ARPES}(d)]; a sharp peak smoothly and monotonically evolves to a broader peak, 
albeit this broadening is more pronounced in the theoretical curves than in experiment.

In a long-range ordered nematic phase, the Fermi surface is increasingly deformed as one moves away from the nodes towards the antinodes [see Fig.~\ref{fig:spectral}(a)].
A related anisotropy characterizes the glassy nematic state, even in the presence of a superconducting gap.
Constant-energy cuts of $A(\bfk, E)$ for a fixed $E$ well above the superconducting gap scale ($E = -0.2t \approx -4 \Delta_0 $) shown in Fig.~\ref{fig:spectral}(b) vividly capture the contrast between the nodal and antinodal regions.
The spectral function is relatively sharp in the nodal region and significantly broadened in the antinodal region.

The corresponding anisotropy is clearly reflected in the imaginary part of the electron self-energy, i.e. the inverse lifetime of the quasiparticles, extracted from the calculated spectral function in the normal (non-superconducting) state of the nematic glass, as shown in Fig.~\ref{fig:spectral}(c).
Moreover, results for two different correlation lengths $\xi_{\mathrm{nem}}=2a$ and $4a$ appear almost identical in their angular dependence; this confirms that our principal qualitative results are robust, despite the short correlation lengths we have assumed for computational simplicity.
Indeed, we find that the exact self-energy extracted from our simulation is qualiatively similar to the self-energy  computed in Born approximation: 
\begin{align}
  \Sigma(\bfk, \omega) 
    &= \int \frac{\mathrm{d}^{2} p}{(2\pi)^2} 
        |g(\bfk, \bfp)|^2 G(\bfp, \omega)  \Gamma(\bfp - \bfk)
\label{eqn:born}
\end{align}
where $g(\bfk, \bfp) \equiv [\tilde{f}(\bfk) + \tilde{f}(\bfp)]/2$,
as shown  in Fig.~\ref{fig:spectral}d.

The energy dependence of the LDOS calculated for the glassy nematic configuration Figs.~\ref{fig:realspace}(a) and \ref{fig:realspace}(b) 
is shown in Fig.~\ref{fig:LDOS}(b);  it exhibits qualitative resemblance to the corresponding experimental data shown in Fig.~\ref{fig:LDOS}(a). The spatial average $\bar{n}(E)$ indicated as the black line in Fig.~\ref{fig:LDOS}(b) has a V shape expected of a uniform $d$-wave superconductor. 
However, the standard deviation $\Delta n(E)$ represented by the shaded region in the same figure grows with energy, and is large at energies larger than and comparable to $\Delta_0=0.055t$. Another way to appreciate the  
``low energy and high energy'' dichotomy is to look at the spatial map of the normalized LDOS $n(\bfx, E) / \bar{n}(E)$
at different energies. 
There is a clear contrast between the relative homogeneity evident in the map at low energy shown in Fig.~\ref{fig:realspace}(c) ($E=0$), and the inhomogeneity of the same map at a higher energy shown in Fig.~\ref{fig:realspace}(d) ($E=-0.2t\sim -4\Delta_0$).
[Note that  quantitative comparison between the experimental results in Fig.~\ref{fig:LDOS}(a) and theory requires some care; for computational purposes (as discussed previously) we have taken a value of $\Delta_0=0.055t$ that is larger than the observed value in experiment.]
%}

We now turn to the optical conductivity, whose temperature and frequency dependences show  trends that are  shared across different material families of underdoped cuprates [see Figs.~\ref{fig:opticon}(a) and (b)].
Well above $T_c$, the real part of the complex conductivity $\sigma_1(\omega)$ is a monotonically decreasing function of $\omega$, as expected of a metallic state.
As the temperature is lowered, the response at $\omega$ below a certain frequency $\omega_{\mathrm{peak}}$ is increasingly suppressed, and $\sigma_1(\omega)$ evolves into a superposition of a sharp peak at $\omega=0$ and a broad peak at $\omega \sim \omega_{\mathrm{peak}}$.
Remarkably, the optical conductivity calculated within our model shows  similar qualitative behavior. 
In the model, the persistence of an increasingly sharp peak at a non-zero energy is a consequence of pair formation.
More importantly, the remaining sharp Drude-like peak with width that tends to zero as $T\to 0$ at small $\omega$ is a manifestation of the coherence of the near-nodal quasiparticles that are largely unscattered in the glassy nematic.

The role of the nematic cold-spot in the optical response can best be seen by comparing the case of the nematic glass in Fig.~\ref{fig:opticon}(c) with the case of point-like scattering in Fig.~\ref{fig:opticon}(d).
When the nodal quasiparticles are scattered by the random potential, there remains a residual density of states at $\omega=0$ even deep in the superconducting phase. As a result, a finite width Drude-like peak persists even as $T\to 0$.

The observed evolution of $\sigma_1$  from a  ``Drude-like'' form at high temperatures to a superposition of a sharp peak at $\omega=0$ and a broad peak at $\omega \sim \omega_{\rm peak}$ is remarkably reproduced by the glassy nematic model. 
However, in the experiments, the crossover between the two forms onsets at  the pseudogap temperature scale $\sim T^*$ well above superconducting $T_c$ while the corresponding crossover onsets at the calculated (mean-field) superconducting $T_c$ in our model. 
Notably, the experimental $\sigma_1(\omega)$ marches through $T_c$ without much notice of it. It is as if d-wave  gap with nodes onsets at $T^*$, with nodal quasiparticles that are largely unscattered as they would be in the presence of glassy nematic order.
We will further discuss constraints on models for $\sigma_1(\omega)$ at temperatures $T_c<T<T^*$ in the next section.
The qualitative similarity between the measured spectra and those calculated from our glassy nematic model in the superconducting state at
$T<T_c$ is not subject to this caveat; it is a robust result of the cold-spots, although the energy scale of the broader peak in experiment is larger than $2\Delta_0$.

\section{Other forms of glassy order}

\begin{figure}
\subfigure[\label{fig:edc-dffcdw}]{\includegraphics[height=2in]{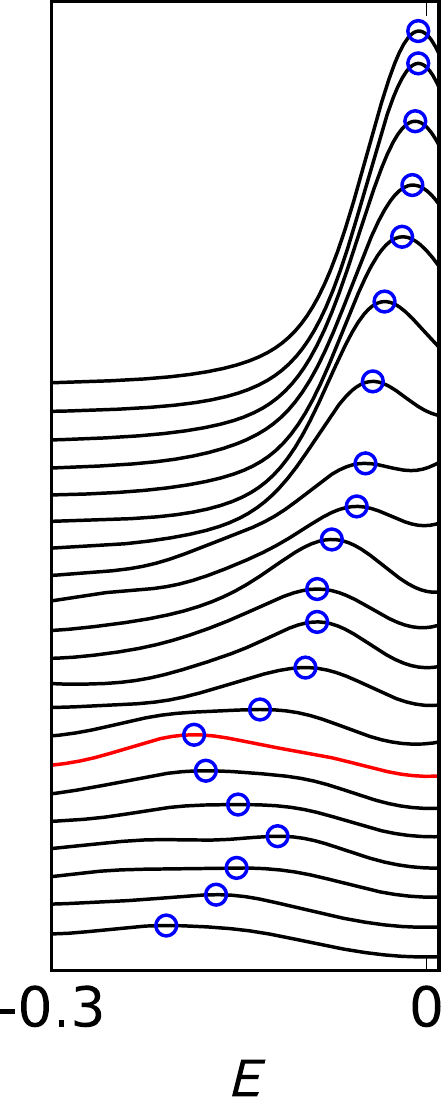}}\qquad
\subfigure[\label{fig:selfenergy-simulation-dffcdw}]{\raisebox{3mm}{\includegraphics[height=1.5in]{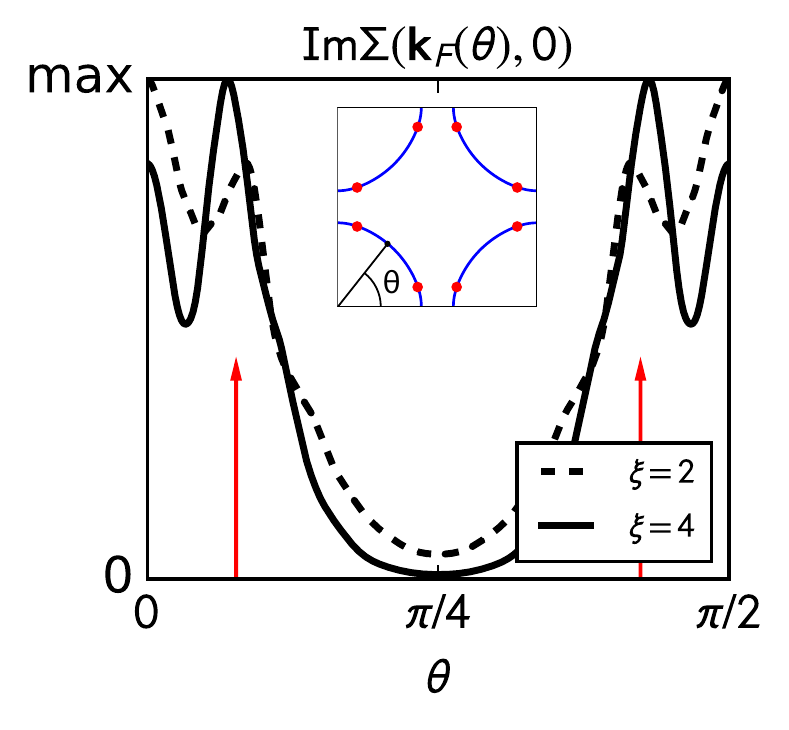}}}
\caption{\label{fig:cdw}
(a) EDCs along the Fermi surface [path 3 of Fig.~\ref{fig:ARPES}(a)], calculated in the superconducting state in the presence of glassy charge-density-wave with a $d$-wave form factor and $\sqrt{\Gamma_{\mathrm{cdw}}} = 0.2t$ and $\xi_{\mathrm{cdw}} = 4a$, multiplied by Fermi function.
The wavevector $\bfQ = (2\pi/a)(0.196, 0)$ is chosen to be a shortest vector connecting the intersection of Fermi surface and the magnetic Brillouin zone boundary, ($|k_x \pm k_y| = \pi$).
Blue circles mark the maxima of each curve, and the red curve indicates the location of a hot-spot momentum.
The angular dependence of the self-energy in the normal state with various values of $\xi_{\mathrm{cdw}}$ is shown in (b).
The hot-spot momenta are indicated by red arrows (red dots in the inset).
}
\end{figure}

We have carried out similar (although less extensive) calculations for various other forms of assumed glassy order.  We comment here briefly on certain aspects of these results.

\subsection{Superconducting $d$-form factor CDW glass} 
While the $d$-wave form factor in the case of a nematic glass is dictated by symmetry, for a CDW in which the ordering vector itself breaks the $C_4$ symmetry of the underlying crystal, the $d$-wave form factor is not symmetry dictated.
Any CDW will thus necessarily have both $s$-wave and $d$-wave components;  conversely, a dominantly $d$-wave form factor presumably reflects some feature of the microscopic physics (the ``mechanism'') which produces the CDW.
Not surprisingly, results obtained for a CDW glass with an assumed $d$-wave form factor share many qualitative features with those obtained for a nematic glass, as these arise from the assumed form factor directly.
The major differences between the two situations concern the existence of ``hot-spots'' on the Fermi surface in the case of the $d$-form-factor CDW glass.
Hot spots refer to the points on the pristine Fermi surface  which are spanned by the CDW ordering vector -- these are the points where, in a weak coupling analysis of CDW order, the effects of the CDW are expected to be most vivid.

The EDC of a $d$-form-factor CDW glass as a function of position along the Fermi surface is shown in Fig.~\ref{fig:cdw}(a) to be compared to Figs.~\ref{fig:ARPES}(d) and \ref{fig:ARPESTheory}(c).
Although we have taken the CDW correlation length in our calculations to be quite short, $\xi_{\mathrm{cdw}}=4a$, (comparable to the CDW wavelength) the existence of a hot-spot is clearly seen in the non-monotonic evolution of the spectral function along the Fermi surface.
This is in sharp contrast with the lack of any such feature in Fig.~\ref{fig:ARPES}(d).
The hot-spots are also visible in the electron self-energy of the normal state along the Fermi surface, as shown in Fig.~\ref{fig:cdw}(b); the hot-spot appears more sharply for longer $\xi_{\mathrm{cdw}}$.

We have not explicitly explored the effects of glassy~\cite{russo_random_2016} ``$d$-density-wave (DDW) order''~\cite{chakravarty_hidden_2001}, because time-reversal-symmetry-breaking required for DDW brings with it additional issues of  modeling.
Nevertheless, since it also has a $d$-wave form factor, we expect that much of the nodal-antinodal dichotomy we have found would also apply to this form of ordering in the superconducting phase.

\begin{figure}
\begin{minipage}{0.48\columnwidth}
\quad\subfigure[]{\includegraphics[height=1.2in]{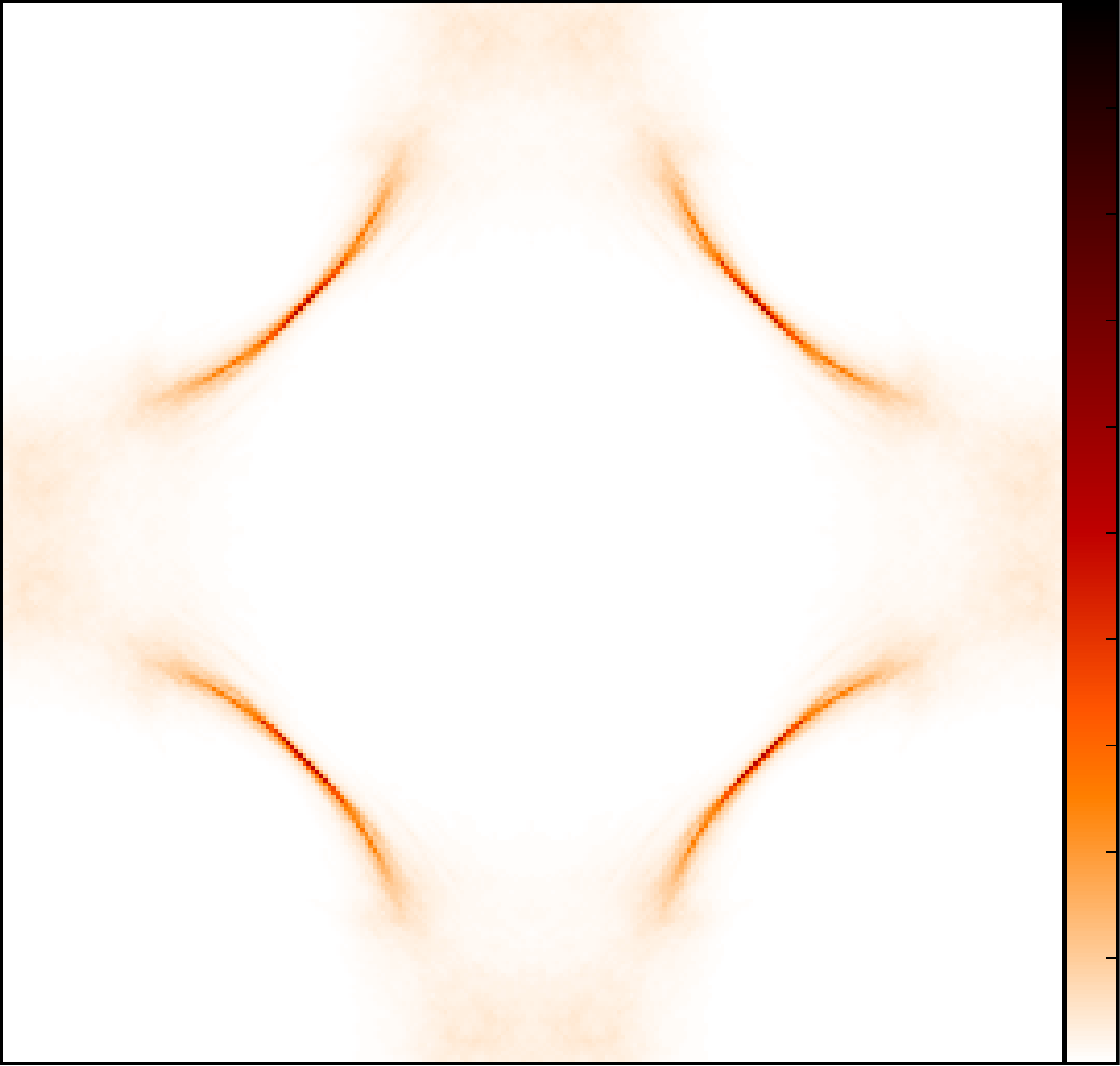}}\\
\subfigure[]{\includegraphics[height=1.2in]{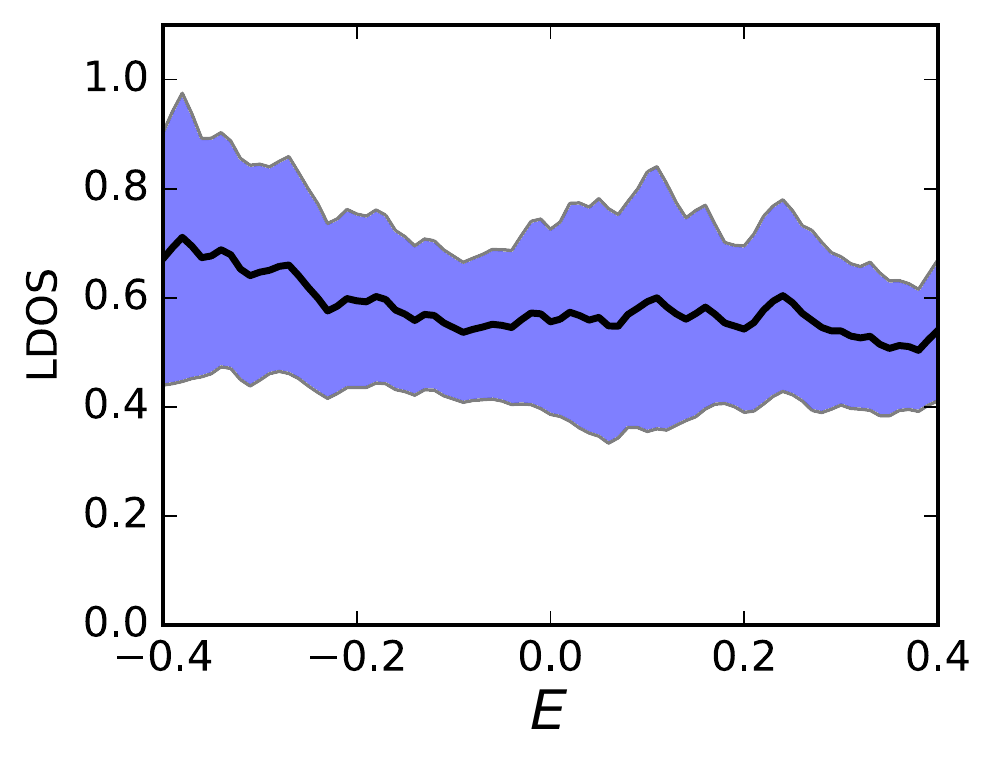}}\\
\subfigure[]{\includegraphics[height=1.3in]{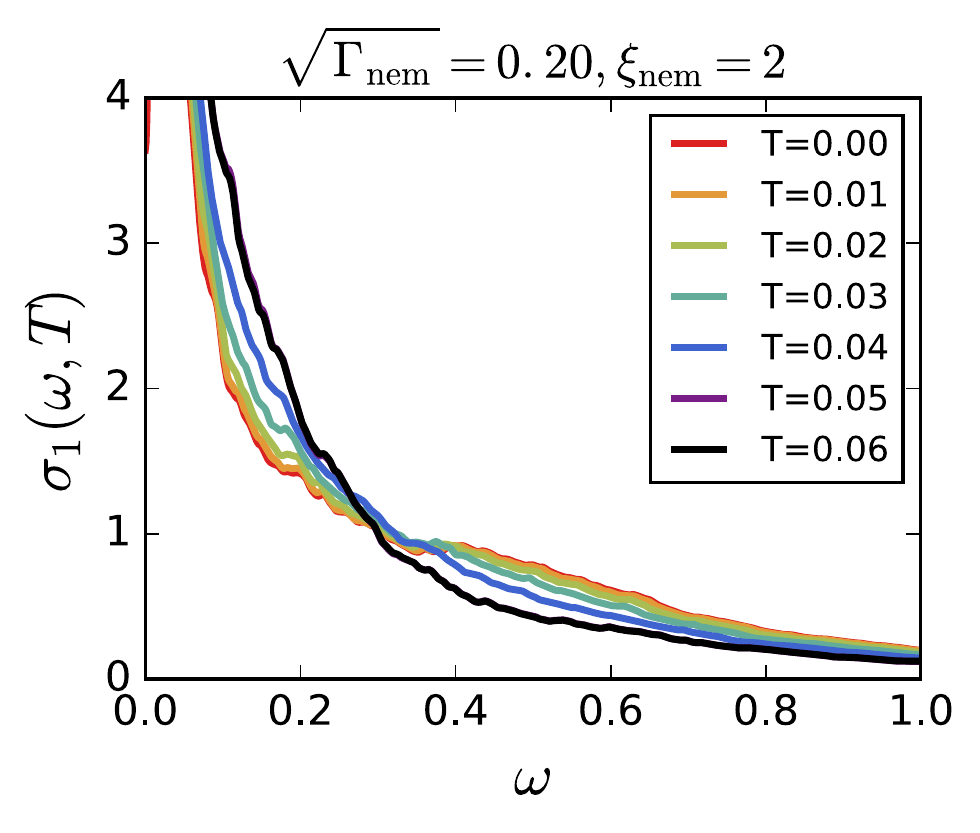}}
\end{minipage}
\vline
\begin{minipage}{0.48\columnwidth}
\quad\subfigure[]{\includegraphics[height=1.2in]{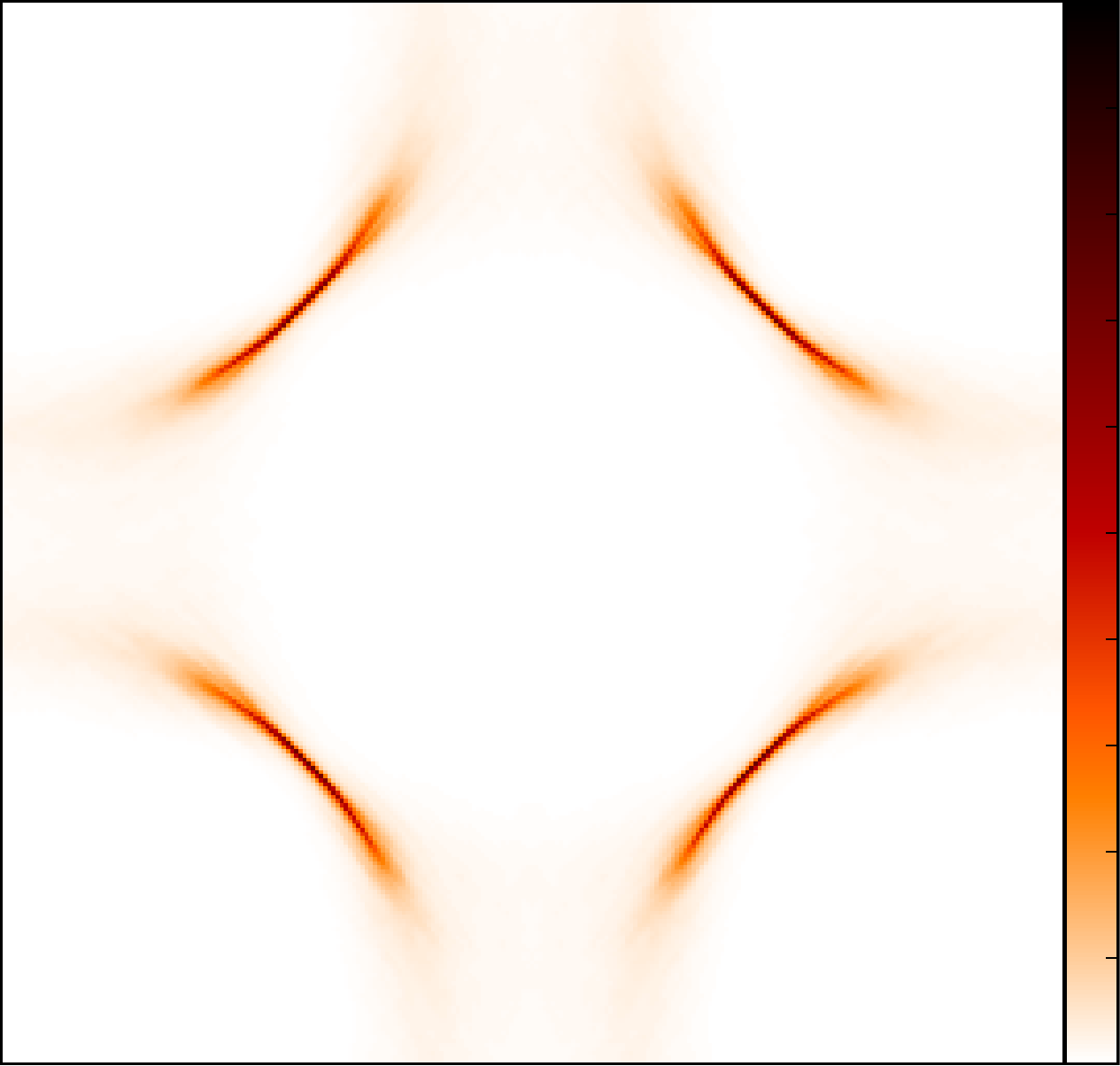}}\\
\subfigure[]{\includegraphics[height=1.2in]{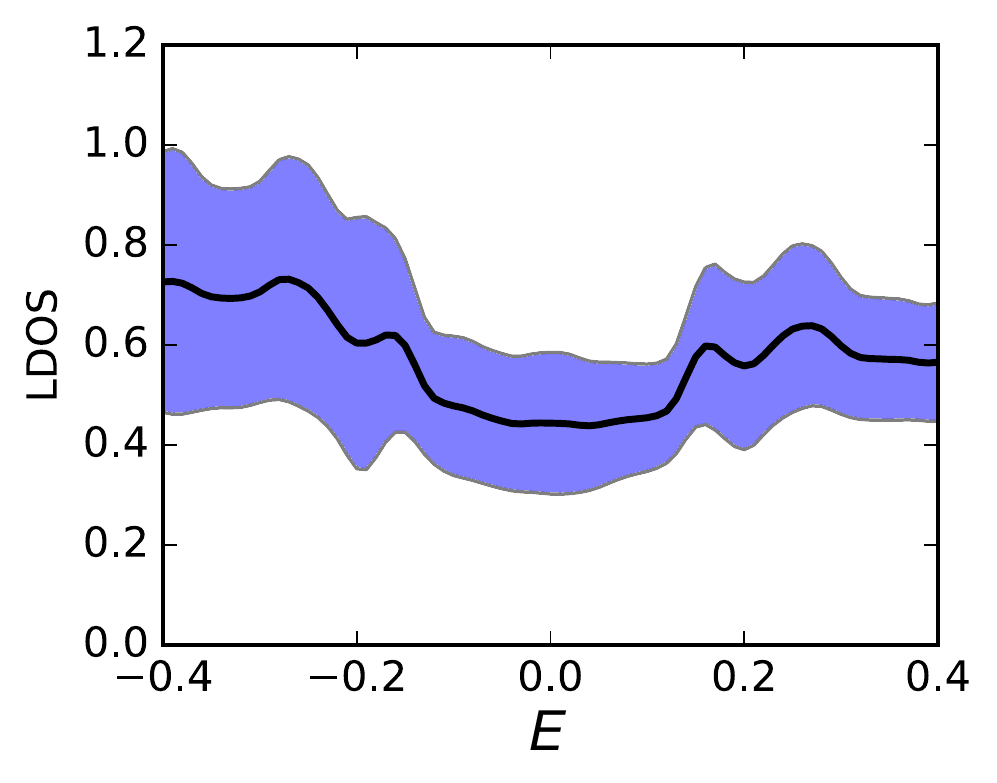}}\\
\subfigure[]{\includegraphics[height=1.3in]{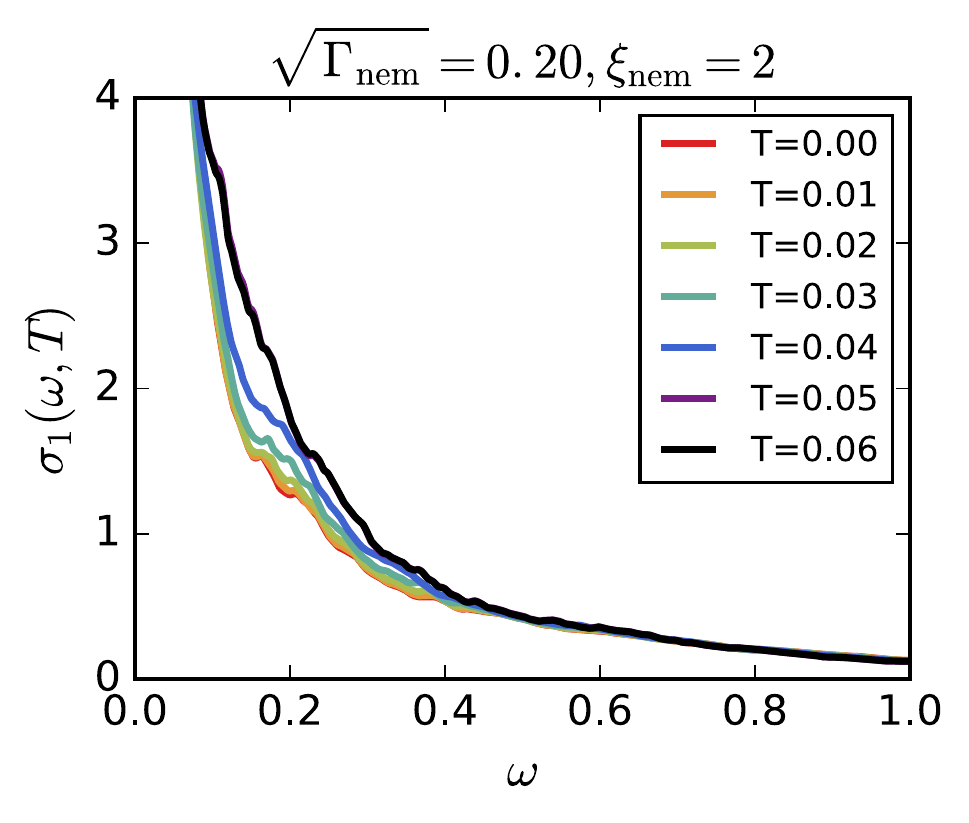}}
\end{minipage}
\caption{\label{fig:otherpossibilities}
Various spectroscopic observables for a nematic glass with coexisting uniform $d$-density-wave order (DDW) (a-c) and with phase disordered (locally $d$-wave) superconductivity (PDdSC) (d-f).
The DDW and PDdSC are set non-self-consistently by Eqs.~\eqref{eq:ddw} and \eqref{eq:pddsc}, with no additional superconducting order.
In the PDdSC,  phase disordering of the superconducting state is represented by incorporating two pinned vortices and two anti-vortices.
(a-c) Spectral function at the Fermi level, local density of states, and optical conductivity with DDW.
(d-f) Spectral function at the Fermi level, local density of states, and optical conductivity with PDdSC.
}
\end{figure}

\subsection{Optical conductivity for other models of antinodal gap}
The fact that the ``two-peak'' structure (a sharp peak at $\omega=0$ and a broad peak at $\omega=\omega_{\rm peak}$) that is well-captured by our superconducting glassy nematic model below $T_c$ persists above $T_c$ in experiments is troublesome.
Our result relies on two essential ingredients of the model to obtain the ``two-peak'' structure:
(1) a d-wave superconducting gap and (2) a d-wave form factor scattering. 
Given the apparent absurdity of assuming the persistence of a superconducting gap far above $T_c$, we 
consider two additional scenarios of a nematic glass with antinodal gaps to seek alternative explanations of the ``two-peak'' structure above $T_c$.

First we consider an ordered $d$-density-wave (DDW)~\cite{schulz_fermisurface_1989,affleck_large_1988,kotliar_resonating_1988,ivanov_staggeredvorticity_2000,nayak_densitywave_2000,chakravarty_hidden_2001} coexisting with glassy nematicity.
We represent the DDW as a contribution to $V$ with 
\begin{align}
  \varphi(x, y) &=i \Delta_{\text{ddw}} (-1)^{x + y}
\label{eq:ddw}
\end{align}
with Eq.~\eqref{eq:coupling} extended to complex field:
\begin{align}
  V(\bfx, \bfy)
    &=
     \frac{1}{2} f(\bfx-\bfy) \left[ \varphi(\bfx) + \varphi^{*}(\bfy) \right]
\end{align}
where $f(\bfr)$ is a $d$-wave form factor.
$\Delta_{\text{ddw}}$ is chosen to be $\Delta_{\text{ddw}} = \Delta_{\text{ddw}}^{0} \sqrt{1-(T/T_c^0)^2}$ non-self-consistently.
To compare with the superconducting state, we chose  $T_{c}^{0} = 0.05t$, and also $\Delta_{\text{ddw}}^0=0.05t$.
This opens a gap at the antinodes, as the Fermi level spectral function in Fig.~\ref{fig:otherpossibilities}a shows.
Nevertheless, the density of states remains finite [as indicated by the finite length of ``arc'' in Fig.~\ref{fig:otherpossibilities}(a), and also Fig.~\ref{fig:otherpossibilities}(b)], leading to the Drude-like peak at $T=0$ in Fig.~\ref{fig:otherpossibilities}(c).
Not surprisingly, DDW order alone is insufficient to account for the nature of the experimentally observed gapping below $T_c$;  even above $T_c$, it does not give as good an account of the structure of the optical conductivity as does the (apparently absurd) assumption of a persistent superconducting gap.

Another way to introduce an antinodal gap is in a model of a ``phase-disordered'' $d$-wave superconductor (PDdSC) with broken time-reversal symmetry.
We introduce minimal phase disorder by incorporating vortices at positions  $(0,0)$ and $(L_x / 2, L_y / 2)$ and antivortices at positions $(L_x/2, 0)$ and $(0, L_y/2)$.
We thus non-self-consistently choose $\Delta_{\bfx\bfy} = f(\bfx-\bfy) \Delta(\frac{\bfx+\bfy}{2})$   to be a product of $d$-wave form factor $f(\bfr) = \delta_{\bfr, \pm\hat{x}} - \delta_{\bfr,\pm \hat{y}}$ and Jacobi theta functions:
\begin{align}
  \Delta(\bfx)
   &=
    \Delta_0 \sqrt{1-\left(\frac{T}{T_c^0}\right)^2}
    \hat{\sigma}_W \left( z; 0 \right) 
    \hat{\sigma}_W \left( z; \frac{\pi}{2}(1+i) \right)
\nonumber\\
   &\qquad\times
    \hat{\sigma}_W^* \left( z;   \frac{\pi}{2} \right)
    \hat{\sigma}_W^* \left( z; i \frac{\pi}{2} \right)
\label{eq:pddsc}
\end{align}
where $z \equiv \frac{x}{L_x} + i \frac{y}{L_y}$, and $\hat{\sigma}_W$ is defined as
\begin{align}
\hat{\sigma}_{W}(z; z_0) &\equiv \sigma_{W}(z; z_0) / | \sigma_{W}(z; z_0) |
\\
\sigma_{W}(z; z_0)
  &\equiv
e^{-\frac{\pi}{2}\left( (z-z_0)^2 -2 z_0^* z \right)}
     \vartheta_3 \left( z - z_0  | i \right) .
\end{align}
Again we choose $T_{c}^{0} = 0.05t$, and $\Delta_{0} = 0.05t$.
Figures~\ref{fig:otherpossibilities}(d)--\ref{fig:otherpossibilities}(f) show the resulting spectra.
The spectral function shows that, while the antinodal excitations become gapped, a large portion of the Fermi surface still survives as noted by \textcite{berg_evolution_2007}.
This Fermi arc leads to a finite density of states in the limit $\omega \rightarrow 0$. In fact the density of states at low energies has a rather flat energy dependence with a suppressed but finite magnitude [Fig.~\ref{fig:otherpossibilities}(e)].
Correspondingly, we find that the optical conductivity does not show any suppression of the low energy spectral weight [Fig.~\ref{fig:otherpossibilities}(f)].
Again, this does not greatly resemble the experimental results for $T>T_c$.
Instead, as with the DDW, the finite density of states at the Fermi level leads to a Drude-like peak at $\omega=0$.
Among the possibilities that we have considered, only the model with $d$-wave superconductivity on top of $d$-form-factor scattering with cold-spots qualitatively reproduce the experimentally measured $\sigma_{1}(\omega)$ for temperatures $T<T^*$.

\section{Conclusion}

In summary, we showed that the consonance between the cold-spots of a glassy nematic and the gap nodes of  a $d$--wave superconductor 
 can account for the most salient ``anomalous'' features of the spectroscopic measurements on the cuprates we have studied.
It is natural in a glassy nematic superconductor that the nodal quasiparticles are long-lived, while away from the nodes, the quasiparticles are strongly perturbed by the local nematic order.
This provides a simple explanation for the nodal-antinodal dichotomy observed by ARPES, and the strongly energy dependent heterogeneity observed by STM. 
Furthermore we found 
striking similarity between the temperature evolution and low temperature form of optical conductivities between our model and experiments. Nevertheless, the fact the   ``two-peak'' structure of the optical response only occurs below $T_c$ within our model, while it persists up to $T^*$ in experiments, implies that fluctuational effects beyond those we have considered must be included in a complete theory of the pseudogap state.

Implicit in the above is the assumption that other sources of quasiparticle scattering -- those associated with point-like ($s$ form factor) disorder or with CDW ordering (either with $s$ or $d$ form factor) -- are relatively weak.
Specifically, as was pointed out previously~\cite{jamei_inferring_2006}, substantial scattering by point-like disorder can be ruled out directly from the experimentally observed sharp V-shape and relative homogeneity of the lowest energy portion of the LDOS.
Above, we have further shown that significant scattering by a CDW with a substantial correlation length -- even one with a $d$ form factor -- can be ruled out on the basis of the lack of any ``hot-spot'' in the observed ARPES spectrum.
Since both point-like disorder potentials and short-range CDW order have been directly imaged in the same sort of BSCCO samples we have used as the basis of these inferences, this raises the issue of why they are so weakly coupled to the low energy quasiparticles~\cite{garg_strong_2008}.

Our results point to interesting future directions. Firstly, a smoking-gun test of our conclusions would be to repeat the spectroscopic measurements on samples under uniaxial strain.
We would predict the ``anomalous features'' to diminish as uniaxial strain detwins nematic domains.
It is also plausible that the response of glassy nematicity to in-field magnetic field may introduce field-dependence of the anisotropic life-time.

\acknowledgements

The authors thank Joseph Orenstein, Ruihua He, and Abhay Pasupathy for useful discussions and comments.
This work was supported by the U.S. Department of Energy, Office of Basic Energy Sciences, Division of Materials Science and Engineering under Award DE-SC0010313 (K.L. and E.-A.K.) and by Department of Energy under grant number DE-AC02-76SF00515 (S.A.K.).

\bibliography{nematicglass}

\end{document}